# Economically viable CO₂ electroreduction embedded within ethylene oxide manufacturing

Magda H. Barecka,*[a] Joel W. Ager [b,c,d] and Alexei A. Lapkin [a,e]

[a.] Cambridge Centre for Advanced Research and Education in Singapore, CARES Ltd. 1 CREATE Way, CREATE Tower #05-05, 138602 Singapore; email: magda.barecka@cares.cam.ac.uk

[b.] Berkeley Educational Alliance for Research in Singapore (BEARS), Ltd., 1 CREATE Way, 138602, Singapore

[c.] Department of Materials Science and Engineering, University of California at Berkeley, Berkeley, California 94720, USA

[d.] Materials Sciences Division and Joint Center for Artificial Photosynthesis, Lawrence Berkeley National Laboratory, Berkeley, California 94720, United States.

[e.] Department of Chemical Engineering and Biotechnology, University of Cambridge, Cambridge CB3 0AS, UK

## Abstract

Electrochemical conversion of CO2 (CO2R) into fuels and chemicals can both reduce CO2 emissions and allow for clean manufacturing in the scenario of significant expansion of renewable power generation. However, large-scale process deployment is currently limited by unfavourable process economics resulting from significant up- and down-stream costs for obtaining pure CO2, separation of reaction products and increased logistical effort. We have discovered a method for economically viable recycling of waste CO2 that addresses these challenges. Our approach is based on integration of a CO2R unit into an existing manufacturing process: ethylene oxide (EO) production, which emits CO2 as a by-product. The standard EO process separates waste CO2 from gas stream, hence the substrate for electroreduction is available at an EO plant at no additional cost. CO2 can be converted into an ethylene-rich stream and recycled on-site back to the EO reactor, which uses ethylene as a raw material, and also the anode product (oxygen) can be simultaneously valorized for the EO production reaction. If powered by a renewable electricity source, the process will significantly (ca. 80%) reduce the CO2 emissions of an EO manufacturing plant. A sensitivity analysis shows that the recycling approach can be economically viable in the short term and that its payback time could be as low as 1-2 years in the regions with higher carbon taxes and/or with access to low-cost electricity sources.

## 1. Introduction

The chemical industry is a large consumer of oil, gas and coal, and is responsible for ca. 15% of industrial $CO_2$ emissions[1]. Consequently, there is a need for significant innovation in this sector in order to achieve necessary reduction targets such as the 2050 goal of carbon neutrality set by the Intergovernmental Panel on Climate Change (IPCC)[2]. Conversion of waste $CO_2$ back to raw materials is a promising path to minimize emissions and reduce the use of non-renewable resources for production, while responding at the same time to the need for chemical products. In this context, this study is motivated by recent reports that $CO_2$ can be electrochemically converted to ethylene at or near the selectivity and conversion rates required for a technologically feasible process[3,4]. We also note significant progress in scaling up $CO_2$ reduction (CO2R) reactors[5–8]. CO2R technology is rapidly approaching the state of commercial application, as exemplified by the introduction by Opus 12 of a 5 kW unit based on the industrially established reactor design for water splitting[9], and other small-scale commercial projects by e.g. Siemens and Evonik, Dioxide Materials, CERT Toronto, Skyre and Mantra Energy Alternative Ltd[10]. In light of these advances, it is timely to consider if it is possible to bring this technology to economically viable operation on the scale of bulk chemicals manufacturing.

Catalyst selectivity, low solubility of $CO_2$, and catalyst stability during extended operation under industrially relevant current densities are known barriers to commercialization and, consequently, are the focus of many research efforts[11–17]. The general process techno-economics are being intensively discussed in literature, targeting identification of most promising products and configurations of electrochemical processes[18–22]. However, from a large-scale deployment perspective, operation of plant-scale CO2R as a stand-alone system will also require additional up- and down-stream processing steps, increased logistic efforts, which contribute significantly to investment and operational costs.

To overcome this limitation, we consider here an alternative strategy for CO2R deployment: on-site $CO_2$ recycling and integration of the electroreduction unit into an existing chemical production process. Waste $CO_2$ produced on site can be directly used for conversion to raw materials and recycled back to the production unit. This approach creates synergies between the emerging CO2R technology and the established chemical industry and minimizes required separation steps and reduces logistical efforts. Furthermore, instead of competing with mature methods for bulk chemicals production, we propose here a CO2R-based retrofit of an existing manufacturing process. Such an approach must (1) significantly reduce $CO_2$ emissions of the chemical plant, (2) ensure a short payback time for CO2R reactor and (3) match the production scale of the chemical. We will show that ethylene oxide is a favourable target for our approach, as all three criteria can be met.

To implement CO2R technology on industrially relevant scale, a high purity $CO_2$ source is required. The majority of experimental studies report operation with a concentrated $CO_2$ stream as a feedstock, hence costly separation of highly diluted $CO_2$ from flue gas is needed, resulting in optimistic $CO_2$ prices of ca. 40\$/ton[22]. The challenge of $CO_2$ capture costs is being addressed by the research community, targeting the use of CO2R units directly with flue gases or air as the $CO_2$ source, though functional systems are currently not ready for implementation. Hence, we considered the opportunity of connecting a CO2R reactor to a production process that already generates a high-purity $CO_2$ waste stream, consequently the raw material for CO2R reactor can be available at no additional cost.

Furthermore, the investment and operational costs related to the CO2R product separation must be carefully considered. Despite of tremendous improvement in catalysts selectivity[23,24,25], only few currently available systems are able to produce a single reaction product with >90% selectivity, and several by-products are found in the post-reactor stream. The feedstock conversion ($CO_2$) rarely exceeds 50%[26]; therefore, significant amount of $CO_2$ is present in the product stream.

Separation costs were highlighted by Jouny et al.[22], who reported a techno-economic analysis for large-scale CO2R units for manufacture of a variety of bulk chemicals. Based on the results of short-cut modelling of ethanol manufacturing with 0.03\$/kWh electricity, even a simplified two-step product separation would account for 40% of capital and 15% of operating costs. A deeper insight into the feasibility of complex flowsheets for $CO_2$-based chemicals production was considered in a life cycle assessment by Greenblatt et al.[27], identifying that a minimum of five separation steps would be needed to recover CO2R products of commercially attractive purity. Furthermore, the operation of described up and downstream sections would include additional logistic efforts, which was determined by Yuan et al.[28] to be one of the critical factors for large-scale deployment of CCU. Thus, we conclude it could be advantageous to look for alternative approach for CO2R operation other than directly making a commodity chemical in a stand-alone production plant.

Due to significant economic limitations in CO2R industrial application, valorisation of the reaction products on the anode side is also being explored, as exemplified by coupling CO2R with oxidation of organics, instead of water oxidation[29,30]. Therefore, while designing a CO2R reactor embedded into an existing chemical production plant, we considered a process for which we can valorise both the cathode and anode outputs as a means of obtaining further processing cost reductions.

We conclude from this overview of prior art, that there is significant potential for increasing the technology's profitability by simultaneous reducing the costs of $CO_2$ capture and of products separation and by implementing the on-site use of the cathode and anode side CO2R products. These costs can potentially be minimized by system integration of $CO_2$ conversion, as concluded by Pérez-Fortes and Tzimas[31] in their industrial study of chemical conversion of $CO_2$ to methanol and formic acid. As outlined above, an attractive target process for CO2R integration should already generate a high purity $CO_2$ waste stream and consume a mixture of products coming from $CO_2$ electro-conversion. In this scenario, a multistep separation chain can be drastically reduced, as $CO_2$ conversion products can be recycled back to the main production process as a mixture.

EO manufacturing (Fig. 1) is an excellent example of a production process into which CO2R can be embedded. Ethylene oxide (EO) is one of the most commonly used commodity chemical with an annual production rate of over 34.5 Mt[32]. It is a key raw material for manufacturing ethylene glycol and specialty chemicals and is also widely used as a disinfectant[33]. The molecule is primarily produced by ethylene epoxidation on silver catalysts[34]. The conversion to EO (R1, Fig.1) proceeds with up to 90% selectivity[35] with complete oxidation of ethylene to $CO_2$ (R2, R3 overall) as a competing process. As a result, ca. 0.4 tons of $CO_2$ are emitted per each ton of EO produced, resulting in emissions of up to 180 kt $CO_2$ annually from a single plant[33]. Production of EO is ranked as one of the most $CO_2$-emitting industries in the chemical sector, along with steam reforming of methane and synthesis of ammonia[2]. Hence, reduction of $CO_2$ emissions from this process can meaningfully contribute to achievement of sustainability goals of the IPCC.

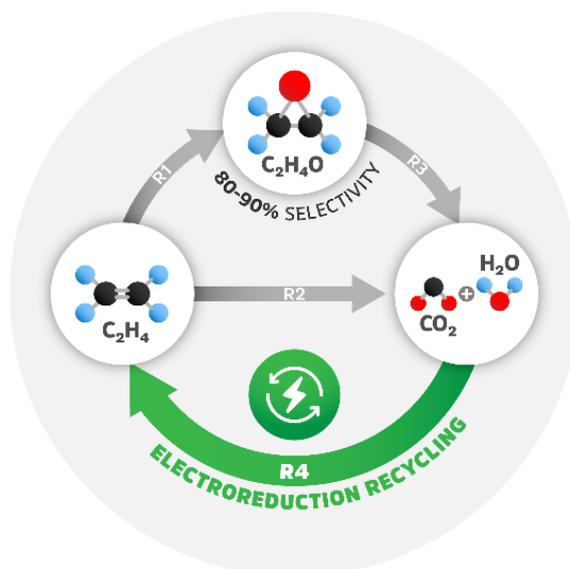

Fig. 1. Proposed pathway for $CO_2$ recycling (R4) within EO manufacture (R1-R3).

Here we report the discovered opportunity to recycle the waste $CO_2$ in the EO process, specifically to regenerate ethylene, the feedstock for the reaction, R4. If the electricity used to drive this process can be sourced renewably, e.g. from wind or solar, $CO_2$ recycling could reduce the overall carbon footprint of the process and enable to store the power of intermittent renewable energy sources in chemicals [36–39]. While the process design and economic analysis of stand-alone $CO_2$ electroreduction plants were already addressed in several papers discussed above, this is the first example of a proposal of integration of CO2R into a well-established bulk chemical production process.

Within this paper, we first describe a base-case EO process using an already optimized variant of the industrial benchmark EO production process[40]. We then demonstrate that CO2R can be embedded into this production process and develop a conceptual flowsheet for integration of $CO_2$ electroreduction (CO2R) into EO plants, based on experimental data for the CO2R reactor performance[24]. The scale-up and cost calculations are derived from the model reported by Jouny et al.[22] for a similar industrial scale stand-alone CO2R unit. These systems can also leverage existing infrastructure and development for PEM water electrolysis[41].

Detailed assumptions for the modelling of the electrochemical reactor are described in Section 4, whereas Section 5 gives a summary of models used for assessment of the capital and operational costs of the integrated design. Next, Section 6 discusses the environment benefit of the integration of the $CO_2$ recycling process and its potential to reduce $CO_2$ emissions from EO plants. Lastly, through sensitivity analysis based on prices of ethylene, electricity, and $CO_2$ taxes we determine the economic scenarios favourable for technology implementation, define techno-economic targets for CO2R units and review the deployment potential for different size of EO plants. Our findings confirm the potential of the recycling process to reduce EO plant $CO_2$ emissions while renewable energies are being integrated, at least partially, into the electricity mix. Further, we identify a number of economic scenarios demonstrating that investment into the recycling process can return in less than 2 years, depending on the $CO_2$ regulatory environment. Benchmarking against other electrolysis processes used in chemical industries (like for e.g. Chlor-Alkali production) we discuss that the scale of CO2R technology required for EO plant integration is feasible, hence the recycling process has a potential for rapid industrial introduction.

| Operating conditions | |
|---|---|
| Reactor residence time | 1.9 s |
| Temperature | 202-210 °C |
| Feed pressure | 10 bar |
| Component | Mole fraction |
| Water | 0.010 |
| Ethylene | 0.389 |
| Carbon dioxide | 0.050 |
| Oxygen | 0.069 |
| Methane | 0.348 |
| Argon | 0.134 |

Table 1. Typical EO manufacturing operating conditions and composition of the reactor inlet stream40. Excess quantities of ethylene and methane as a make-up chemical are used due to safety constraints; argon is introduced as impurity and partially accumulated for dilution purposes.

## 2. Ethylene Oxide base-case production plant

The process flow sheet for the state-of-the-art EO technology (defined here as the base-case[40], Fig.2) consists of EO production (Section 1), EO removal and purification (Sections 2 and 3) and $CO_2$ (by-product) removal (Section 4). The conversion reaction is direct catalytic oxidation of ethylene (frequently also called ethylene epoxidation) in the gas phase, using pure oxygen as the oxidant and silver-based catalysts[33]. In addition to the desired partial oxidation of ethylene to EO, two side reactions occur that limit the EO yield: the complete oxidation to carbon dioxide ($CO_2$) and water and the consecutive combustion of EO to $CO_2$ (Fig.1, R2 and R3). The reaction takes place at elevated temperature (>200 °C) and pressure (>10 bar). Due to safety constraints (high risk of explosion), the substrates (ethylene used in excess and oxygen) are diluted with methane and limited quantities of reaction by-products and impurities which are partially accumulated in the reactor. Table 1 lists typical feed composition and reaction conditions.

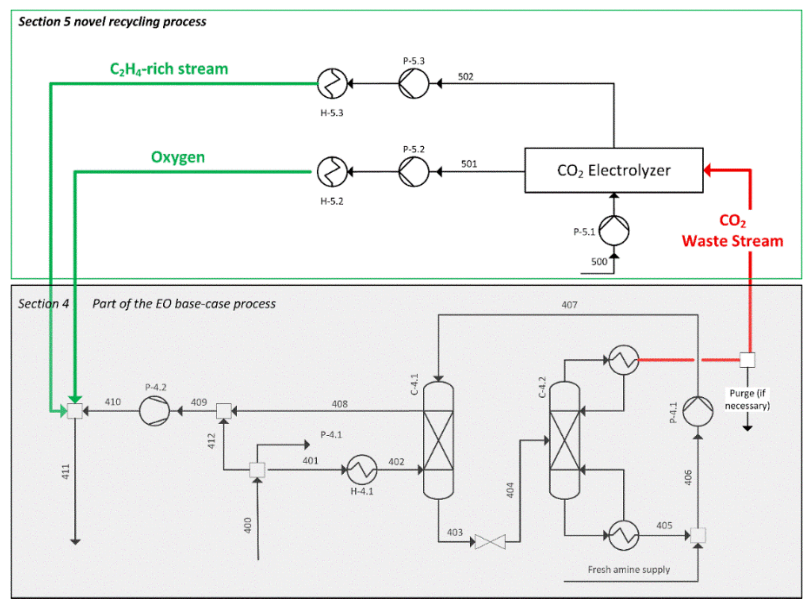

Fig. 3. A conceptual flowsheet of integration of the CO2 recycling process (Section 5) into EO manufacture (Base-Case Design Sections 1-3 remain unchanged).

The product mixture is sent to the downstream treatment section, which separates EO by means of high-pressure absorption with water (Section 2), after which EO is desorbed at low pressure in a stripper and further purified by distillation (Section 3). In the $CO_2$ removal section, the side product is scrubbed by chemical absorption in a hot potassium carbonate ($K_2CO_3$) solution (Section 4). This is the most common, but not the only reagent used. Only partial removal of $CO_2$ is necessary (quantity generated by side-reaction), as some $CO_2$ is recycled back to the reactor and used for dilution purposes. $CO_2$ is desorbed at atmospheric pressure and usually released into the atmosphere. The post-reaction stream depleted in EO and partially in $CO_2$ is mixed with a fresh supply of ethylene, oxygen and make-up methane, preheated and recycled back to the reactor.

| Reaction | Faradaic efficiency |
|---|---|
| $CO_2 + 2H^+ + 2e^- \leftrightarrow CO + H_2O$ | 3.8% |
| $2CO_2 + 12H^+ + 12e^- \leftrightarrow C_2H_4 + 4H_2O$ | 88.7% |
| $2H^+ + 2e^- \leftrightarrow H_2$ | 7.5% |

Table 2. Electrochemical reactions taking placed in the modelled electrolyzer and Faradaic efficiencies reported24, accounting for the possible measurement error.

## 3. CO2 recycling: plant integration concept

For a typical EO plant, the cost of ethylene corresponds to over 90% of the overall manufacturing costs, hence generation of the side product $CO_2$ limits profitability. To date, the majority of efforts aimed at reducing the amount of $CO_2$ produced during ethylene epoxidation have focused on catalyst improvement and addition of inhibitors[42–45]. These strategies have increased the selectivity towards EO and therefore have reduced $CO_2$ generation; nevertheless, the highest selectivity achieved so far is around 90% and generation of the by-product seems inevitable[33]. Therefore, we proposed a retrofit based on valorisation of this waste stream.

Within the proposed novel flowsheet (Fig. 3, Flowsheet Section 5) the waste $CO_2$ stream, separated in the base-case EO process design through absorption and desorption, is sent directly to the $CO_2$ electrolyzer, fed also with recycled catholyte. On the cathode side of the $CO_2$ conversion reactor, an ethylene-rich stream is produced (stream nr 502), including also a minor amount of carbon oxide and hydrogen as well as unreacted $CO_2$. The quantity of unreacted $CO_2$ results from feed conversion, which is a key process operational parameter discussed in the reactor modelling (Section 4.1). In the EO plants, a specific quantity of $CO_2$ is used as feed to the EO reactor ($CO_2$ is used for dilution purposes in usual fraction of 5% of the feed stream), therefore unreacted $CO_2$ can be recycled back to the EO reactor without violating the operational constraints. Other electroreduction products such as carbon oxide and hydrogen are obtained in minor quantities and once diluted with the remaining flow of the fresh ethylene send to the EO reactor, will be present in ppm levels, similarly to other impurities fed to the reactor (see SI for details). Existing impurities as well as additional quantity of carbon oxide and hydrogen can be removed by a purge stream already integrated in EO plants (see Fig.2). On the anode side, the only compound recovered and send back to the EO reactor is oxygen (50% valorisation assumed).

Flowsheet Sections 1-3 of the base-case design for EO manufacturing are not affected by the $CO_2$ electroreduction integration, and hence are not displayed in Fig. 3.

## 4. Process modelling

Due to the lack of applicable process simulators for the electroreduction unit, dedicated short-cut models were used to design each processing step and relate our findings to the scale of EO manufacturing. In the base-case EO plant[40], 103 mol/s of waste $CO_2$ are continuously emitted, corresponding to the amount of $CO_2$ generated as EO production reaction by-product. The composition of the feed $CO_2$ stream is considered as reported for the base-case EO design[40]. Minor impurities accounting for ca. 0.3% of inlet stream are not modelled (See Process Streams Summary Table 4.)

The design of the CO2R unit is based on an electrode system with an excellent selectivity towards the desired ethylene product (FE = 88.7% after the consideration of the potential experimental error)[24] at the current density of 300 mA/cm$^2$. Similar ethylene selectivity have been reported for different copper-based electrode systems[46,47,48,49,50,51] with the newest findings reaching FE 65-75%[4,23] at current densities as high as >1000 mA/cm$^2$, and long-term stability testing demonstrated[3,46,52]. The observed system of electrochemical reactions, along with the determined Faradaic efficiencies (Eq. 1) are given in Table 2. Based on these experimental results, product flows are calculated.

$$\varepsilon_{Far} = \frac{z \cdot n \cdot F}{Q} \tag{1}$$

$\varepsilon_{Far}$ is Faradaic efficiency, z is the number of electrons transferred, n is the amount of product, F is Faraday constant, Q is the total charge passed.

Precise design and sizing of the CO2R unit is not currently possible due to the lack of commercially developed $CO_2$ electrolysers of sufficient capacity. Thus, to provide a complete process concept and to evaluate its economic potential, engineering approximations derived from commercialized technologies were used for scale-up calculations. Key process parameters (Table 3) were extracted from experimental reports and compared to two commercially established electroreduction processes: water splitting and Chlor-Alkali $Cl_2$ production[53,54]. This benchmarking enables us to verify if the assumed parameters are feasible to implement in the current state-of-the-art electroreduction reactor technology. For our modelling we fixed the current density at 300 mA/cm$^2$, though leveraging from chemical electroreduction processes discussed above, there is a potential to increase the current density at least 2x, decreasing the necessary electrolyzer area. The need for high current operation (even up to 1000-2500 mA/cm$^2$) was recently discussed[19,55] and researcher's attention is being brought to testing $CO_2$ reduction catalyst systems only at current densities relevant for industrial scale deployment[56]. Cell voltage (2 V) was extrapolated from the experimental report[24] as a value potential to be achieved in an optimized system; we anticipate also that an industrial scale system would operate under minimized potential losses as opposed to lab-scale systems. Therefore, our full cell voltage value is slightly lower than in other recent experimental demonstrations[3,24].

| Design parameters | Value used for modelling | Industrial benchmarks[53,54] |
|---|---|---|
| Current density | 300 mA/cm$^2$ | Water splitting: 75-2000 mA/cm$^2$ <br> Chlor Alkali: 700 mA/cm$^2$ |
| Cell voltage | 2 V | Water splitting. 1.75-2.4 V |

Table 3. Selected CO2 electrolyzer parameters vs. industrial electroreduction benchmarks.

Though of extreme importance[26], one-pass conversion of $CO_2$ is infrequently reported, as single pass reactors with small electrode surface are used for lab-scale experiments. In work aimed specifically at optimizing one-pass efficiency, values between 22.5% to 68% have been reported[57,58,59] and recent TEAs usually use values around 50% as reference[19,22,21], though the conversion is likely to be significantly improved upon scale-up efforts. Hence, we evaluate the environmental and economic performance for conversion varying between 50-100% and present different operational modes for the continuous process.

In case of conversion close to 100%, whole waste $CO_2$ stream is converted to ethylene-rich stream and directly recycled back to the EO reactor. If lower conversions are achieved (e.g. 50%), unreacted $CO_2$ is recycled back to EO reactor in the ethylene-rich stream. Whereas this recycling does not impose any limitation to the EO reactor, already operating with certain $CO_2$ quantity, removal of the complete amount of $CO_2$ generated as a by-product must be ensured to avoid the compound build-up. Consequently, the stream of $CO_2$ separated in the absorption column needs to be slightly increased, since instead of venting the whole $CO_2$ stream, some $CO_2$ is recycled back to the reactor together with the CO2R products. The cost of the altered operation of the absorption-desorption is not considered in the following design, as the section is responsible for a negligible part of the operational costs for the EO process (0.1%). Hence, any change in these costs has an insignificant impact. Additionally, a purge of $CO_2$ before the electrolyzer is used to remove a part of separated $CO_2$.

Key electroreduction reactor parameters such as the necessary electrode area, current, power and process streams (Table 4) are calculated based on the assumptions listed above, applying the model proposed by Jouny et al.[22]. In contrast, our process design is based on an experimentally demonstrated reaction selectivity for a complex system and not assumed on the best-case values for the single product $CO_2$ conversion. We also consider the effect of the potential measurements error and extracted the worst-case scenario data from the experiments (see Supplementary Information).

| Component | Waste CO$_2$ (mol/s) & catholyte | Anode-side output (mol/s) | Cathode-side output (mol/s) |
|---|---|---|---|
| CH$_4$ | 0.7 | | 0.7 |
| C$_2$H$_4$ | 2.1 | | 47.7 |
| CO | | | 11.7 |
| H$_2$ | | | 23.2 |
| H$_2$O | 114.4 | | |
| CO$_2$ | 103.0 | | |
| O$_2$ | | 154.3 | |

Table 4. A summary of main components for process streams for the integrated CO2 conversion section (assumed 100% conversion).

Several properties of the reactor system are strongly dependent on temperature. Intriguingly, Dufek et al.[60] have found that temperature rise from 18 to 70 °C increased the overall cell efficiency by 18% for the liquid phase operation, whereas Lee et al.[61] observed doubling of the partial current density at the same voltage as a result of elevated temperature of a gas-phase electrolyzer. The operation in elevated temperatures might be favourable as an increasing share of the energy requirement for the reaction can be covered by thermal energy, being intensively explored in high-temperature electrolysis design[62]. In order to rigorously evaluate the heat balance of the large scale CO2R reactor, modelling approach reported for another electrochemical process (Alumina reduction) was used[63]. The total enthalpy required for a conversion reaction is converted first to the voltage equivalent (Eq. 2). The difference between the cell voltage (2 V) and the enthalpy voltage equivalent represents total potential losses throughout the electrochemical activation, Ohmic heating effect and other minor sources. These potential losses correspond to the maximum heat produced by CO2R reactor (Eq. 3).

$$E_{\Delta H^0} = \Delta H^0(x)/(nF) \tag{2}$$

$$Q = E_{Cell} - E_{\Delta H^0} \tag{3}$$

where $E_{\Delta H^0}$ is the voltage equivalent of enthalpy, $\Delta H^0$ is standard enthalpy of reaction, x is Faradaic efficiency, n is the number of electrons transferred, F is Faraday constant, E is the cell voltage, Q is dissipated heat.

Despite the potential benefit of high-temperature operation, the total heat generated in the CO$_2$ rector was assumed to be removed to follow the economically worst–case scenario (high cooling requirement). Consequently, our model assumes an isothermal operation. The investment cost for the heat management system is already included in the reference stack cost. The operational cost of heat removal with the cost indicator for cooling extracted from EO base-case design process was evaluated and proves to have a minimal impact on the operational cost. Notably, due to the dependence of Ohmic losses on geometry of a cell, Ohmic heating effect can be engineered to certain extent, allowing for optimal heat management. Therefore, a deeper understanding of thermal processes occurring on the electrode surface, and rigorous optimization of the reaction would provide further margin for improvement of the system' performance. The recovered heat stream could be also potentially integrated and used on-site in the EO plant.

Before the recycling of the cathode output, the stream is send to a phase split unit. Both cathode and anode side gaseous outputs are subsequently compressed to meet the EO reactor' feed requirements (10 bar, 202 °C). Energy consumption for compressors (Eq. 4-5), are calculated using engineering approximations for similar scale systems[64].

$$W_c = \frac{n_f}{\eta_c} \cdot \left(\frac{\gamma - 1}{\gamma}\right) \cdot R \cdot T_{in} \cdot \left[\left(\frac{P}{p}\right)^{\frac{\gamma-1}{\gamma}} - 1\right] \cdot 10^{-6} \tag{4}$$

$$\frac{T_{out}}{T_{in}} = \left(\frac{P}{p}\right)^{\frac{\gamma-1}{\gamma}} \tag{5}$$

Wc is the power required by the compressor, ηc is compression efficiency, γ is adiabatic expansion coefficient, R is the universal gas constant, Tin/Tout is the ratio of inlet to outlet temperatures.

## 5. Capital and operational costs calculation

### 5.1. Capital costs

Rigorous verification of the technology feasibility requires a detailed analysis of capital costs. The electrolyzer capital cost calculation is performed with the DOE H2A analysis[65], implemented into the $CO_2$ electroreduction cost model[22]. The benchmark electrolyzer costs 250.25 $/kW and operates at 0.175 A/cm$^2$; 1.75 V (Norsk Hydro HPE Atmospheric Type No. 5040 alkaline electrolyzer). This cost was recalculated per surface area, calculated for 100% conversion (with installation factor of 1.2) and multiplied by the necessary area of the $CO_2$ electrolyzer. The sum of the costs of the supporting components (balance of the plant) is assumed to be 35% of the total cost.

The cost of compressor and heat exchangers units are leveraged from economic coefficients for a comparable system[64]. The overall investment cost comprises the costs of the electrolyzer unit, balance of the plant, and the connected compressors and heat exchangers. The costs of site preparation, engineering and design, up-front permitting cost are out of scope of this analysis as these contributors are due to significant location and business-model dependent variations and hence cannot be taken into account with sufficient accuracy.

### 5.2. Operational costs and income calculations

Key operating costs account for energy consumption by the electrolyzer unit and supporting equipment, whereas the added value generated by the integrated flowsheet is roughly equal to the cost of ethylene and oxygen which are recycled back to EO production reactor and minimize the need for fresh reactants. For the sake of comparability with the base-case design process, the cost indicators for electricity and ethylene are taken from EO base-case design[40]. The cost of water for the reaction of 0.0054 $/gal is extracted from a recent TEA analysis[22]. Based on the process modelling results and the determined energy requirements, the overall operational cost is calculated. Further, maintenance costs of 2.5% of capital costs per year are added.

Table 5 gives a complete overview of all cost calculations for the scenario of 100% $CO_2$ conversion. In case of a lower conversion, the electrode area is proportionally decreased, thus reducing both the investment costs and the savings throughout the generated ethylene stream. Consequently, payback time evaluated as the overall investment cost divided by the early profit is not conversion-scale dependent is equal to 1 year for the base-case cost indicators scenario.

| Capital costs | | |
|---|---|---|
| *Electrolyzer unit* | | |
| H2A Base Cost | 250.25 | $/kW ($2010) |
| Base case Electrolyzer Voltage | 1.75 | V |
| Installed Cost for $CO_2$ elctrolyzer (300 mA/cm$^2$, 2V) | 919.7 | $/m$^2$ |
| Electrolyzer Maintenance cost: | 2.5 | % capital per year |
| Total Electrolyzer cost: | 18.3 | M$ |
| Balance of Plant | 9.83 | M$ |
| *Additional equipment* | | |
| Compressors (M$) | 0.008 | M$ |
| **Operating costs** | | |
| Cost of electricity | 0.00124 | $/MJ |
| Power required for electrolyzer | 4.255 | M$/y |
| Power required for compressors | 0.12 | MW |
| Cost of compressors operation | 0.0043 | M$/y |
| Cost of water | 0.00003 | $/mol |
| Water consumption | 0.190 | M$/y |
| Maintenance cost (M$/y) | 0.46 | M$/y |
| Overall operating costs (M$/y) | 4.8 | M$/y |
| Yearly profit (M$/y) | 28.34 | M$/y |
| Payback time (y) | 1.0 | y |

Table 5. A summary of capital and operational costs for the CO2 recycling process assuming 100% CO2 conversion. The payback time evaluated as the overall investment cost divided by the early profit is not conversion-scale dependent.

## 6. Potential for reduction of CO2 emissions

Throughout the complete conversion of waste $CO_2$ back to ethylene, the EO production plant with the integrated recycling process does not vent $CO_2$ to the atmosphere, as opposed to the current state-of-the-art EO flowsheet. In the case of a lower conversion being achieved (e.g. 50%), the direct emissions are reduced proportionally to $CO_2$ conversion. Nevertheless, from the perspective of the overall $CO_2$ cycle, both direct and indirect $CO_2$ emission sources must be taken into account. Hence, we evaluate the $CO_2$ emissions balance for the new recycling section considering the additional consumption of utilities (electricity). Fig. 4 shows the potential for $CO_2$ emissions reduction for recycling with 50% and 100% conversions, powered by different energy sources. Additionally, on a global scale, recycling results also in savings through the on-site generation of the main product - ethylene, which avoids production of this amount of ethylene from petrochemical feedstocks (not quantified in our $CO_2$ emissions balance). $CO_2$ emission coefficients for different sources of electricity were extracted from the most detailed industrial reports available[66,67].

The recycling process offers an excellent opportunity to reduce the impact of direct $CO_2$ emissions by ca. 80%, provided that low-carbon power such as solar photovoltaics, geothermal or hydropower is used. Such sources need to be at least partly integrated in the grid to allow for optimal $CO_2$ balancing, or be available on-site. For example, electricity coming from gas combined cycle need to be mixed with ca. 60% of hydro-originating power to ensure zero overall $CO_2$ balance for the recycling section. This emerging technology both levels the use of intermittent electricity from renewables and provides economic incentive for renewable electricity industry by establishing new long-term fixed-price contracts. Our findings on $CO_2$ emissions reduction potential are in line with previously published recommendations to introduce carbon capture and utilization within the chemical industry[68,69].

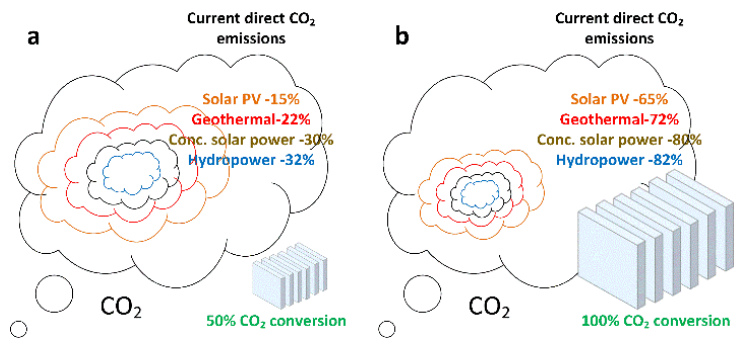

Fig. 4. Illustration of the potential for reduction of CO2 emissions in EO plants for different sources of energy used for electroreduction under a) 50% and b) 100% CO2 conversion.

## 7. Cost sensitivity analysis
### 7.1. Investment cost

Following the demonstrated potential of the recycling process to reduce $CO_2$ emissions, we discuss different economic scenarios favourable for the technology deployment. Firstly we analyse the investment costs, constituting mainly of the CO2R reactor stack cost, with its cost being estimated from the large-scale electroreduction model analysed in the US Department of Energy Hydrogen Analysis (H2A) project [65]. This approach covers a wide range of manufacturing scales and models and has been used as a basis for a number of recent CO2R TEA studies[22,19]. Different scenarios can be extracted from this report[65], resulting in the overall stack cost ranging from 1,000 to 15,000 \$/m$^2$. In the analysis of Jouny et al.[22], the averaged cost indicator for optimized large scale production is being considered. Our analysis targets similar scale, hence, the same benchmark is used in this study.

However, as the electrolyzer stack cost indicator could vary depending on manufacturing scenario, we investigate how these variations influence the total investment cost. Notably, the latter cost is also strong function of the applied current, which determines the total area of the required electrode. The presented CO2R design is based on experimentally proved density of 300 mA/m$^2$, however the parameter could be significantly increased with the development of large scale electrolysers (up to 1000-2500 mA/cm$^2$)[19,55]. Fig. 5 shows that higher electrolyzer stack costs could be balanced by operation under increased current densities, which is being intensively investigated with first reports[4] on operation at over 1000 mA/cm$^2$. Cost-optimization of large-scale units along with high current density operation can secure viable investment costs, supporting large-scale technology deployment.

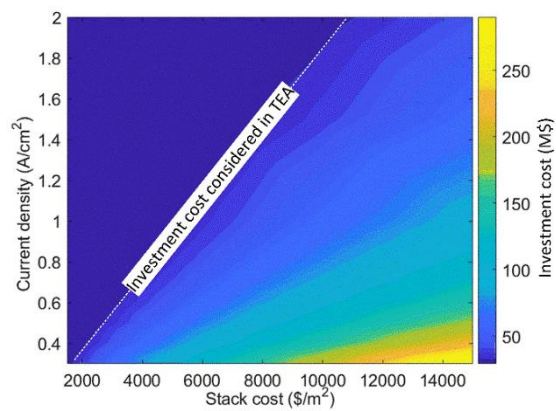

Fig. 5. Sensitivity analysis for investment cost: dependence of the overall investment cost on the cost indicator for the stack unit and the applied current density. White dotted line represents the investment cost considered in this case-study. Results are given for a system designed for maximum (100%) CO2 conversion.

### 7.2. Operational costs

The operational manufacturing cost for the recycling process are supported by low electricity prices, minimizing the operational cost of the electroreduction unit as well as by high ethylene cost which makes CO2R manufacturing competitive. The cost indicators for those industrial utilities are dependent on project location, scale of manufacturing and available electricity mix. Furthermore, electricity price for the industrial use is significantly lower than for the domestic use. The base-case design process for EO production has low cost indicators for both ethylene (0.021 $/mol) and electricity (0.004 $/kWh), reported specifically for EO manufacturing[70] and used in different reports on EO manufacturing optimization[40,71]. We therefore rigorously consider the effect of the grid electricity and ethylene price and the ROI throughout the sensitivity analysis. We do not evaluate the effect of oxygen price changes, as this raw material is in many cases produced on-site from ambient air and its price is not due to significant fluctuations related to the crude oil market dynamics. However, we consider that in response to raising concerns about climate change and as a matter of support for deployment of novel CCU technologies[72], several countries have introduced $CO_2$ emissions taxes ranging between few cents up to $130 per tonne of $CO_2$.

Based on the review of different $CO_2$ restriction policies adapted worldwide[73], we defined four relevant levels of $CO_2$ taxes (zero/low/medium/extreme). For each of the $CO_2$ tax scenarios, we highlight the ROI time (being not $CO_2$ conversion scale dependent) for the current averaged prices of electricity, including renewables (0.03 $/kWh)[74,75,76] and averaged price of ethylene (0.032 $/mol)[22,77]. From this staring point, economically viable scenarios defined by ROI≈3 years and benchmarked by a lower limit for ethylene price and an upper limit for electricity price:

**Zero $CO_2$ tax economies.** Ethylene price>0.025 $/mole or electricity price<0.040 $/kWh.

**Low $CO_2$ tax economies** (taxes up to $15/tonne $CO_2$, for e.g. Singapore, Japan, South Africa). Ethylene price>0.023 $/mole or electricity price<0.042 $/kWh.

**Medium $CO_2$ tax economies** (taxes between $15- $50/tonne $CO_2$, for e.g. Canada, various EU countries). Ethylene price>0.020 $/mole or electricity price<0.045 $/kWh.

**Extreme $CO_2$ tax economies** (taxes between $50- $130/tonne $CO_2$, for e.g. Switzerland, Sweden). Ethylene price>0.013 $/mole or electricity price<0.057 $/kWh.

Fig. 6 presents the outcomes of our sensitivity analysis, where the payback time for the proposed integrated flowsheet is evaluated under different values of key performance factors such as grid electricity and ethylene prices for the determined level of $CO_2$ taxes.

As highlighted throughout our analysis, a promising ROI (<two years) is determined for this technology with current prices of ethylene and electricity and zero $CO_2$ tax. The ROI can be further reduced, provided that at least medium $CO_2$ taxed are to be paid. Numerous countries applying currently middle and high $CO_2$ taxes are planning its step-wise increase in next years, which increases the economic viability of the $CO_2$ recycling process. For the already highly restricted $CO_2$ economies, the recycling scenario is viable for deployment under a wide range of ethylene and electricity prices with currently a very short ROI (<one year). As demonstrated, to balance the life-cycle $CO_2$ emissions, CO2R should operate mainly with renewable or nuclear energy sources. Higher integration of these energy sources into the energy infrastructure is forecasted to reduce electricity production costs down and consequently, increase profitability of $CO_2$ electroreduction.

Lastly, fluctuations in ethylene prices are complex to predict, however due the ongoing depletion of fossil fuels, the increase of the cost of the key raw materials is inevitable, which favours efforts towards increasing the efficiency of the use of raw materials and its recycling by means of dedicated CO2R retrofit. Key cost calculations are summarized in the Supporting Information and can be used to accurately assess the technology costs for different prices of electricity and raw materials.

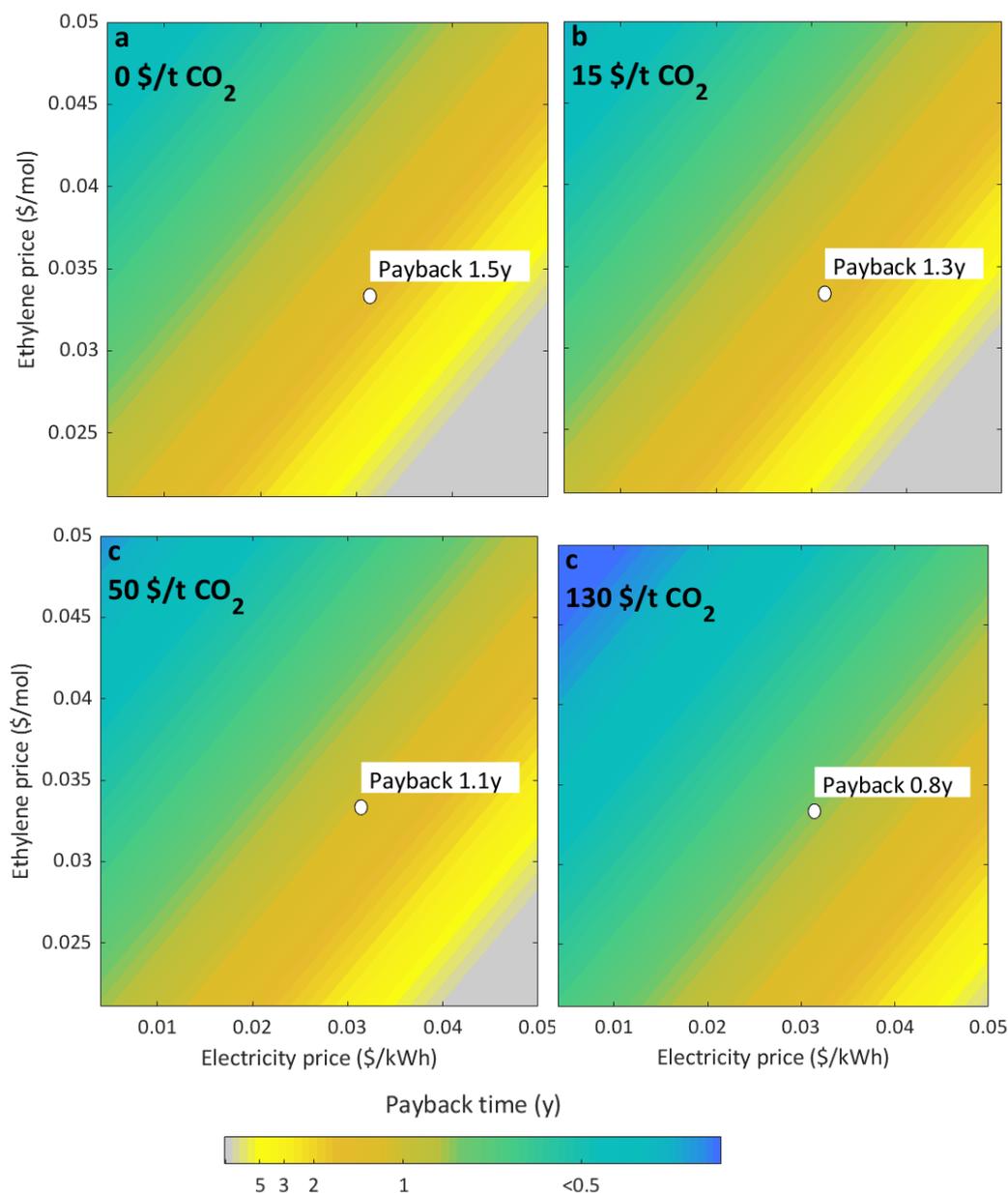

Fig. 6. Relationship between electricity and ethylene prices and investment payback for different CO2 taxes scenario: a) no CO2 tax, b) $15/ton of CO2, b) 50$/ton of CO2, b) 130$/ton of CO2. Grey areas represent inviable scenarios, white dot reflects current electricity and ethylene prices and resulting ROI time.

## 8. Discussion of the results

The presented conceptual design illustrates how to integrate $CO_2$ electroreduction into an established process flowsheet and simultaneously satisfy the EO manufacturing constraints. We showcase that CO2R integration enables to increase the overall mass efficiency of ethylene conversion, being the key bottleneck in EO production plants[40,78] and in many other bulk chemicals production processes[79]. On-site generation of ethylene facilitates logistic efforts and provides a strategy to respond to highly fluctuating ethylene prices.

In contrast to stand-alone CO2R plants, our design does not rely on $CO_2$ separation or purchase cost, as a high-purity $CO_2$ stream is already available at EO plants. In a recent review[80], de Luna et al. highlight the dependence of CO2R on the maturity of carbon capture technologies, hence deployment-wise it is of crucial importance to remove this bottleneck. To further improve CO2R economics, we optimally integrate key CO2R products both on cathode and on anode side and reduce necessary downstream processing operations. The emerging $CO_2$ electroreduction technology is applied on large-scale in the most simplified manner, in an environment matching the need for CO2R feed and products at the same place and time.

As industrial-scale $CO_2$ electroreduction is an emerging technology, the developed process model is based on several assumptions and analogies derived from similar commercialized units. Consequently, successful implementation of the process requires first further stride to ensure long-term operational stability of CO2R units and high selectivity under industrially relevant conversion (>50%). Recently published experimental reports demonstrate significant progress in the field and that the technology benchmarks used in this study (Faradaic efficiency, cell voltage, current density) are achievable on a short-time scale[3,22,81]. Nevertheless, a targeted experimental investigation is necessary for a pilot-plant scale CO2R unit, its sensitivity to minor impurities present in the recycled gases, as well as stable reactor scale up and engineering of thermal effects.

To identify most promising techno-economic scenarios for the deployment of the $CO_2$ recycling process, sensitivity analysis for both investment and operational cost is presented. Throughout the maximum use of existing infrastructure, the investment costs are reduced to mainly the cost of CO2R stack. As the ROI is demonstrated not to be dependent on the achieved conversion, the process has a potential to be first implemented within smaller scale electroreduction reactors and gradually increase the conversion scale in parallel to accumulating the large scale processing experience.

Furthermore, we demonstrate that in case of integrated CO2R units, the dependence of the investment costs on the applied current density is significantly higher than for stand-alone CO2R plants[19]. Therefore, significant reduction of investment costs will be possible if further strides are made to demonstrate selective CO2R operation under current densities higher than considered in this study.

The operational costs are primary a function of the electricity and ethylene prices. From carbon abatement perspective, the recycling process should operate with low-carbon energy sources. Recently renewable electricity is being available at more attractive prices: wind power purchase dropped below 0.02 $/kWh[82], utility-scale photovoltaic solar dropped the cost by almost 80% between 2000 and 2017 and further reduction is foreseen in close future[83]. Availability of such low cost sources of carbon-neutral energy enables our process to operate in an economically viable way regardless of ethylene price and applied $CO_2$ levies policy. Importantly, as our technology operates as an add-on to the existing production process, the $CO_2$ recycling does not have to operate continuously and can be activated only upon the availability of the renewable energy.

In addition to the economic benefit of integration of renewable resources, the constant and long-term electricity requirements created by the proposed $CO_2$ recycling process support establishing new fixed-price contracts and penetration of renewable energy into the electricity supply mix[84]. The concept and economics of using renewable electricity for cycling $CO_2$ in the chemical industry are increasingly being discussed, with different configurations of the future combined electricity-chemical manufacturing system being presented[18,80,84–86]. Large-scale deployment of power-to-chemicals concept is being considered in the long-term action catalogue of German Enquete Commission, identifying electrochemical processes as key contributors to sustainable chemistry[87]. Our recycling process can meaningfully contribute to this strategic transition within the chemical industry sector.

To connect our results to the scale of worldwide EO manufacturing, we calculated the required electrode area for different capacities of EO plants (from 5 to ca 400 kta of EO annually), varying from our case study. Intriguingly, Fig. 7 shows that we can relate the scale of electrodes necessary for such production to the existing industrial examples of other electrolysis unit. A PEM electrolyzer of 167 m$^2$ was reported[55], which would be roughly the scale necessary to integrate CO2R based recycling for a 5 kta small-scale on demand EO plant. However, majority of EO plants operate on scales of 100 kta and higher. A 345 kta plant such BASF EO unit in Ludwigshafen (Germany) would require over 19,000 m$^2$; this may seem large, though the existing water electrolyzers or Chlor-Alkali Plants do operate at even higher scale (up to 37,500m$^2$). In recent review, Smith et al.[55] analyse the power to fuel system on a scale capable of influencing the energy system and recognize that the largest electro catalytic systems are available are far smaller than required for producing for e.g. 10,000 t of methanol per day, concluding that a significant gap exists between available systems and what is necessary on global scale. Significantly, our design addresses this challenge, as for retrofit purposes we require a smaller electrode area, yet we are able to limit the impact of big-scale chemicals processing by use of CO2R technologies on feasible scales. The experience from proposed CO2R deployment can be leveraged in future to even higher processing scales.

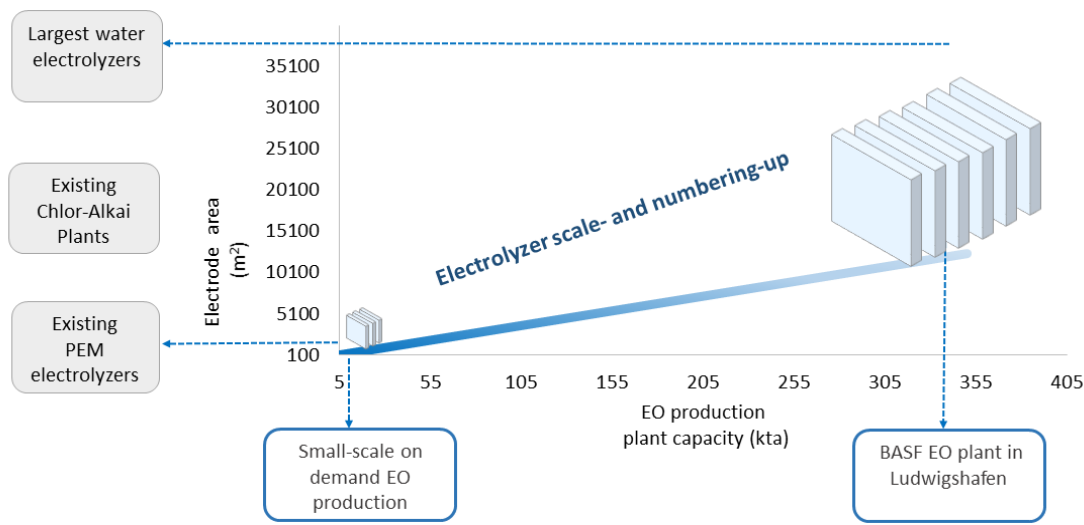

Fig. 7. Perspective of deployment process for EO manufacturing: relationship between EO plant capacity and the required electrode area for the integrated CO2 recycling process with complete CO2 conversion.

## 9. Conclusions

We show that integration of a $CO_2$ electroconversion unit into EO manufacturing can both reduce costs and lower the carbon footprint. As our approach deploys CO2R by means of integration with an existing manufacturing process, it becomes possible to significantly reduce the capital and operational costs of the promising, yet economically challenging $CO_2$ electroconversion technology. As a result, one of the key bottlenecks in the industrial implementation of CO2R can be significantly reduced. A sensitivity analysis incorporating electricity and ethylene prices and considering a number of $CO_2$ tax scenarios shows payback times in the range of five years to as short as one year depending on local conditions. From a sustainability perspective, our process will drastically reduce $CO_2$ emissions from the production of one of the most important bulk chemicals and, further, creates an economic incentive for penetration of renewable energy into the well-established bulk chemicals market.

The proposed method is potentially transformative and deployable over the full range of sizes of EO manufacturing plants, following necessary validation and optimization efforts. We show how to reduce the risk of the application of an emerging technology as we target valorising of a waste stream with minimized interruption to the existing manufacturing process. A significant advantage of this approach is that it does not compete with existing methods of making commodity chemicals but rather provides an overall strategy to retrofit existing processes on a carbon constrained market. This scheme has the potential to be adapted to other integrated processes through exploiting synergies of waste $CO_2$ streams and other process streams, with the emerging new catalytic technologies for conversion of $CO_2$.

## Conflicts of interest

The authors filed patent applications US 62/987,369 and US 63/036,477 for $CO_2$ recycling in different ethylene-based plants, including ethylene oxide, and several recycling process configurations, including impurities removal. The are no other conflicts of interest to declare.

## Acknowledgements

This work was supported by the National Research Foundation (NRF), Prime Minister's Office, Singapore under its Campus for Research Excellence and Technological Enterprise (CREATE) programme through the eCO2EP project operated by the Cambridge Centre for Advanced Research and Education in Singapore (CARES) and the Berkeley Educational Alliance for Research in Singapore (BEARS).

# Notes and references

# Supporting information

Contents

# CO2R Unit Calculations

*Inlet flow rate calculations*

The electrolyser is designed to converted between 50% and 100% of $CO_2$ stream emitted by a typical medium-size EO plant[1].

- Plant capacity: ca. 330 kta for EO production
- $CO_2$ waste stream generated: 103 mol/s (this amount of $CO_2$ is generated as reaction by-product)

In case of low conversion (50%): The complete stream of separated $CO_2$ is sent to the electroreduction unit. Unconverted $CO_2$ is recycled back to the reactor with ethylene and other gases. As small part of $CO_2$ is recycled, the stream of $CO_2$ separated in the absorption column is readjusted to ensure that overall mass balances are met:

$$CO_2 \text{ to be separated} = \frac{CO_2 \text{ from side reaction}}{1 - CO2R \text{ conversion efficiency} \cdot \text{membrane separation efficiency}} \quad (1)$$

$$CO_2 \text{ to be separated} = 103 \frac{mol}{s} / (1 - 0.5 \cdot 0.5) = 137 \frac{mol}{s} \quad (2)$$

From EO manufacturing perspective it is most profitable to convert $CO_2$ into an ethylene-rich stream. Hence, a highly ethylene-selective system was chosen[2].

- Electrode: Cu
- Current density: 300 mA/cm$^2$
- Faradaic efficiencies: values reported[2] were normalized to Faradaic efficiency =100% to account for possible experimental errors and extract the worst-case scenario from the experimental data.

*CO2R Products flow rate calculations:*

$$\text{product flow}(\frac{mol}{s \cdot cm^2}) = \frac{\text{current}(\frac{A}{cm^2}) \cdot \text{Faradaic efficiency}}{\text{nb of electrones transferred} \cdot \text{Faraday constant}(\frac{C}{mol})} \quad (3)$$

*Required electrode area*

Moles of $CO_2$ converted during reaction are calculated stoichiometrically. In order to achieve certain conversion of available $CO_2$, following electrode area is needed:

$$\text{electrode area }(m^2) = \frac{\text{available } CO_2(\frac{mol}{s}) \cdot CO_2 \text{ conversion}}{CO_2 \text{ converted during reaction }(\frac{mol}{s \cdot cm^2}) \cdot 10000(\frac{cm^2}{m^2})} \quad (4)$$

Based on electrode area, flow of each product is calculated. (See Article Table 4).

*Current required for operation (based on main product flow)*

$$\text{current (A)} = \frac{\text{product flow}\left(\frac{mol}{s}\right) \cdot \text{nb of electrones needed} \cdot \text{Faraday constant}\left(\frac{C}{mol}\right)}{\text{Faradaic efficiency}} \quad (5)$$

*Power needed*

$$\text{power} = \text{cell voltage (V)} \cdot \text{total current (A)} \quad (6)$$

## Process streams calculations

*Electrolyzer inlet (gas)*

Composition of this stream is the same as waste $CO_2$ stream reported for the base-case design EO process[1]. The size of the stream is readjusted in order to ensure enough $CO_2$ removal from process gases (see point 1). Minor impurities accounting for ca. 0.3% of inlet stream are not modelled.

*Electrolyzer outlet (gas)*

This stream includes reaction products, minor quantities of compounds introduced with the feed (like e.g. ethylene) and unconverted $CO_2$.

*Ethylene-rich stream recycled back to EO reactor*

The cathode output of the $CO_2$ electroreduction unit will consist also of a minor quantity of carbon oxide and hydrogen, which are usually not used in EO reactors. We simulated what would be the concertation of these compounds when diluted with the fresh ethylene feed to the EO reactor (Table 1). According to the simulation results, carbon oxide and hydrogen would be present in ppm levels which have a potential to be acceptable for the reactor operation. To avoid the compounds build-up, carbon oxide and hydrogen can be removed from recycled gas stream by the existing purge unit, used for removal of other impurities such as e.g. ethane.

| Compound | Mass% |
|---|---|
| $H_2O$ | 0.7% |
| $C_2H_4$ | 41.2% |
| $CO_2$ | 8.3% |
| $C_2H_4O$ | 0 % |
| $O_2$ | 8.4% |
| $CH_4$ | 21.1% |
| Ar | 20.3% |

| $C_2H_6$ | 441 ppm |
|---|---|
| *Post-electrolysis impurities* | |
| CO | 254 ppm |
| $H_2$ | 36 ppm |

Table 1. Simulated composition of the EO reactor feed with possible post-electrolysis impurities.

## Capital cost calculations

*Electrolyzer capital cost & balance of the plant*

Electrolyzer capital costs calculation is based on DOE H2A analysis[3], used previously for cost estimation of big-scale $CO_2$ electroreduction unit by Jouny et al.[4]. The electrolyzer analyzed by DOE costs 250.25 $/kW and operates at 0.175 A/cm²; 1.75 V. This cost was recalculated per surface area (with installation factor of 1.2) and multiplied by the area of the $CO_2$ electrolyzer.

$$CO_2 \; electrolyzer \; cost \qquad (7)$$
$$= refrerence \; cost \left(\frac{\$}{kW}\right) \cdot current \; density \left(\frac{A}{cm^2}\right) \cdot cell \; voltage \; (V)$$
$$\cdot inst.factor \cdot electrolyzer \; area$$

Balance of the plant is assumed to be 35% of the total cost and the stack is 65%.

## Operational costs calculation

*Utilities cost indicators*

Cost of electricity, $C_2H_4$ and $CH_4$ is taken from EO base-case design[1]. Cost of water for reaction 0.0054 $/gal is taken as reported by Jouny et al.[4] and recalculated to $/mol:

$$water \; cost = 0.0054 \frac{\$}{gal} \cdot \frac{1}{3.785} \frac{gal}{l} \cdot \frac{l}{kg} \cdot 0.018 \frac{kg}{mol} = 0.00003 \frac{\$}{mol} \qquad (8)$$

*Heat balance of the electrolyzer*

The total enthalpy required for an electrochemical reaction is converted first to the voltage equivalent. The difference between the cell voltage (2V) and the enthalpy voltage equivalent represents total potential losses throughout the electrochemical activation, Ohmic heating effect and other minor sources. Those potential losses correspond to the maximum heat produced by CO2R reactor. The empirical gas-phase molar-enthalpy data is extracted from NIST databases[5]. Even though the operation in elevated temperatures might be favorable[6,7], the total heat generated in the $CO_2$ rector was assumed to be removed to follow the economically worst–case scenario.

*Maintenance costs*

Maintenance costs of electrolyzer are roughly evaluated as 2.5% of capital costs per year:

$$maintenance\ cost = 2.5\% \cdot capital\ cost \qquad (9)$$

*Overall operating costs*

Overall operating costs are the sum of electricity consumption by the electrolyzer unit, compressors and maintenance costs.

## CO2 emissions balance

To assess whether our process indeed minimizes $CO_2$ emissions, we calculated the $CO_2$ emissions balance for the new processing section (Section 5, Fig. 3), as the difference between the additional utilities consumption (electricity) and $CO_2$ emissions savings. $CO_2$ emission coefficients were extracted from the most detailed industrial reports available[8,9]

$$CO_2\ emissions\ for\ Section\ 5 \qquad (10)$$
$$= -CO_2\ flowrate\ \left(\frac{kg_{CO2}}{s}\right) + electricity\ consumption\ \left(\frac{MJ}{s}\right) \cdot \frac{1\ kWh}{3.6\ MJ}$$
$$\cdot CO_2\ emission\ coeff, el.\left(\frac{kg_{CO2}}{kWh}\right)$$

We focus here on electricity consumption due to its prevailing effect on the environmental impact. Furthermore, other minor source of impact like for e.g. energy use for cooling have a great potential to be further minimized throughout heat integration.

# List of symbols

*Latin symbols*

| | |
|---|---|
| $F$ | Faraday constant (C/mol) |
| $n$ | Number of moles (mol) |
| $n_f$ | Total feed flow rate (kmol/s) |
| $P$ | Feed pressure (bar) |
| $Q$ | Total flow of charge (C) |
| $R$ | Gas constant (J/mol K) |
| $T_f$ | Feed temperature (K) |
| $T_{out}$ | Outlet temperature compressor (K) |
| $W_c$ | Power required for compressor (MW) |
| $z$ | Number of electrons (-) |

*Greek symbols*

| | |
|---|---|
| $\zeta_{Far}$ | Faradaic efficiency (-) |
| $\gamma$ | Adiabatic expansion coefficient (-) |
| $\eta_c$ | Compressor efficiency (-) |

**Supplemental references**